\documentclass[pdflatex,sn-mathphys-num]{sn-jnl}
\usepackage{lipsum}
\usepackage{graphicx}
\usepackage{dcolumn}
\usepackage{bm}
\usepackage{color}
\usepackage{comment}
\usepackage{physics}
\usepackage{blindtext}
\usepackage{svg}
\usepackage{epsfig}

\begin{document}
\title{Thermal quantum information capacity in a topological insulator}

\author*[1]{\fnm{Leonardo A.} \sur{Navarro-Labastida}}
{
\centering
\affil*[1]{Depto. de Sistemas Complejos, Instituto de F\'isica,

Apdo. Postal 20-364, 01000, CDMX, M\'exico}
}

\abstract{Thermal effects in a one-dimensional Su-Schrieffer-Hegger topological insulator are studied. Particularly, we focus on quantum information processing capacity for thermal ensembles. To evaluate quantum information processing an optimized quantum Fisher information is introduced as a quantifier of entanglement and topological phases are calculated by a definition in real space for the electric polarization of mixture states. For the thermal ensemble, there is a relationship between the Fisher metric and the electric polarization in such a way that in the topological region, there is more entanglement, and therefore, creates more robustness and protection in the quantum information against to thermal effects. Moreover, long-range hopping effects are studied and it is found that in this case, the optimized quantum Fisher information captures these topological phase transitions in the limit of low temperature by the formalism in real space. }
\maketitle

\section{Introduction} 
Quantum metrology is in charge of studying the optimization of measurement processes in quantum systems \cite{MODI2012, Victtorio2006}, for example, photonic systems, cold atoms, and superconducting qubits are quite sensible systems \cite{KAI2019, KOSUKE2016}; therefore, finding ways of measuring prepared states in these systems is a critical issue to technology. Entanglement in pure systems is a good way to measure physical properties \cite{Horodecki2009} and determines entirely the behavior of quantum states. However, quantum matter presents interactions between particles and has decoherence and noise that produce complex behaviors of quantum systems. All of this is the reason why it is necessary to find new approximations to measure quantum information processing in macroscopic systems with such characteristics beyond pure entanglement \cite{SZALAY2015, MODI2012}. The study of quantum entanglement that measures non-locality in quantum systems opens a way to understand the entanglement behavior of particles. Quantum information is a more general quantum metric that entanglement and gives more information about the quantum states and quantum geometry of the system. In quantum information theory, quantum discord (QD) is a measure of quantumness between subsystems \cite{MODI2012, Allegra2011}. QD is a type of quantum correlation related to physical effects that do not necessarily involve entanglement \cite{MODI2012,Eberley2006}. \\
\\
An important quantity in quantum metrology is the quantum fisher information (QFI) \cite{Hauke2016, Rezakhani2019}. The classical Fisher information is a way of measuring the amount of information that an observable carries about an unknown parameter of a distribution \cite{HALL2000}. On the other hand, QFI is related to the uncertainty of measuring an observable by its related operator over a complete set of states. The minimization of the quantum information is an optimization process that is related to the lowest bound of entanglement on the systems, while the upper bound of the QFI quantifies the maximal quantum information process capacity \cite{Adesso2014, MODI2012}. The minimization of QFI is called Optimized QFI (OQFI), is related to the minimum eigenvalue \cite{Adesso2014} and tells us how entanglement plays an important role in observables like electron transport or thermal conductivity \cite{Lambert2019,Ugo2017}, besides OQFI characterizes some phase transitions in anisotropic systems \cite{RAN2018}. \\
\\
There are many candidates to make a quantum computer but in general, having stable states and scalability properties is desirable for a good platform in quantum computing. Particularly, one-dimensional (1D) and two-dimensional (2D) topological crystals are systems with highly stable states that are experimentally feasible for example most recently it seems that twisted bilayer graphene is a good candidate because of its strong interactions leading to unconventional superconductivity, coherence localization properties, and quantum metric \cite{Navarro-Naumis2021,Navarro-Naumis2022,rmf2023,rmf2023arxiv,Navarro-Naumis2023,Navarro-Naumis2024}. Recent works find that topological phases are related to a more efficient process of quantum information \cite{YIXIN2010,RAN2018,Ugo2017} and also to qubit transfer and modulation of blocks of information. In many-body systems, entanglement is not enough to characterize the presence of topological phases, the reason why entanglement metrics become more important in complex physical systems to quantify their quantum properties. Here we study thermal effects in the 1D extended Su-Schrieffer-Hegger (SSH) model \cite{CHAO2010}.
\\
\\
The ensemble topological phases are calculated numerically by the extrapolation to mixture states of the Resta phase in real space using periodic boundary conditions \cite{RESTA1998,RESTA2000}. The OQFI is calculated as a measure of entanglement due to thermal effects, furthermore, the uncertainty relation to measure particular states is discussed. 
\section{Model and formalism}
The system under study is the extended SSH Hamiltonian \cite{CHAO2010} this is a tight-binding model of a wire with an alternating single and double hopping. The basis of the wire is constructed by a cell of two types of atoms A and B.
The Hamiltonian in real space can be written as, 
\begin{equation}
\begin{split} 
\hat{H}&= \sum^N_{m}[(v\hat{m}+w_{+}\hat{m}_{+})\otimes\hat{\sigma}_{x}+iw_{-}\hat{m}_{-}\otimes\hat{\sigma}_{y}]
\end{split},
\end{equation}
where the second neighbor hopping is $z$ and the hopping $v$ and $w$ are the first neighbor intra-hopping and inter-hopping interactions in 1D topological wire, respectively. Here $w_{\pm}=w\pm z$ and the projector operators are defined as,
\begin{equation}
\begin{split} 
\hat{m}_{\pm}=\frac{1}{2}(\hat{P}_{m+1,m}\pm\hat{P}_{m,m+1})
\end{split},
\end{equation}
and 
\begin{equation}
\begin{split} 
\hat{m}=\hat{P}_{m,m}
\end{split},
\end{equation}
with $\hat{P}_{m^{\prime},m}=\ket{m^{\prime}}\bra{m}$ and $\hat{\sigma}_{i}$ with $i=x,y,z$ as the Pauli $SU(2)$ matrices. The simple (extended) SSH Hamiltonian occurs when $z=0$ ($z\neq0$).
The basis $\ket{m}$ is the index of the cell number, $\ket{A}$ and $\ket{B}$ are the occupation of the atom type in a cell with the relations 
\begin{equation}
\begin{split} 
\hat{\sigma}_{x}=\ket{B}\bra{A}+\ket{A}\bra{B}
\end{split},
\end{equation}
and 
\begin{equation}
\begin{split} 
\hat{\sigma}_{y}=i(\ket{B}\bra{A}-\ket{A}\bra{B})
\end{split},
\end{equation}

This model defines the essential form for the bipartite system of $2\otimes N$ dimension and sets the delocalization over the topological chain. The eigenfunctions can be written as a superposition of the composite states,

\begin{equation}
\begin{split} 
\ket{m,\alpha}=\ket{m}\otimes\ket{\alpha}\in\mathcal{H}_{site}\otimes\mathcal{H}_{atom}
\end{split},
\end{equation}

here $\ket{m}\in\mathcal{H}_{site}$ is the external dimension or degree of freedom related to the position over the chain, while $\ket{\alpha}\in\mathcal{H}_{atom}$ represent the internal degree of freedom with $\alpha\in{(A,B)}$ related to the type of atom. 
If we introduce an additional hopping to second neighbors characterized by the parameter $z$, this hopping allows the transfer of atoms A to atoms B in two cells and can be represented as the projector $\ket{m+1}\bra{m}\otimes\ket{B}\bra{A}$, therefore, the simple SSH Hamiltonian is a particular case of the one presented in (Eq. 1), that is obtained for $z\neq0$. 

The eigensolution of the Hamiltonian (Eq. 1), $H\ket{\psi_n}=E_{n} \ket{\psi_n}$ can be expressed as a combination of the composite state, $\ket{\psi_n}=\sum^N_{m}(C^A_{m,n}\ket{m,A}+C^B_{m,n}\ket{m,B})$,  where $C^{\alpha}_{m}$ are the amplitude of probability of the particle to be in cell $\ket{m} $ and atom $\ket{\alpha}$. \\

\subsection{Topological phases in a thermal bath} 
In the context of topological systems phases, like the Berry phase are defined for pure states. In ensemble systems, a mixture of topological phases needs to be implemented. For this purpose, we calculate numerically the electrical polarization for an ensemble by the definition of Resta polarization \cite{RESTA2000}-\cite{RESTA1998} as follows,
\begin{equation}
\begin{split}
\boldsymbol{P}=\frac{e}{2\pi}Im\ln{Tr[\rho e^{i\delta\hat{x}}]}
\end{split},
\end{equation}
where $\delta=\frac{2\pi}{Na}$ and $\hat{x}=\sum^N_{m}\hat{x}_m = \sum^N_{m}m[\ket{m,A}+\ket{m,B}]$, with "$e$" the charge of the electron, "$a$" the atomic distance in natural unities, i.e. $e=a=1$ and $\hat{X}=e^{i\delta\hat{x}}$. It follows that for pure states the topological phase reduces to $\gamma_n=Im\ln{[\bra{\psi_n}e^{i\delta\hat{x}}\ket{\psi_n}]}$, so $P_n=\frac{e}{2\pi}\gamma_n$; therefore, the electrical polarization is proportional to the topological phase \cite{RESTA1998,RESTA2000}. Here periodic boundary conditions were used for the numerical calculation of the Resta polarization.

\begin{figure}[hbt!]
\centering
    \includegraphics[width=12cm]{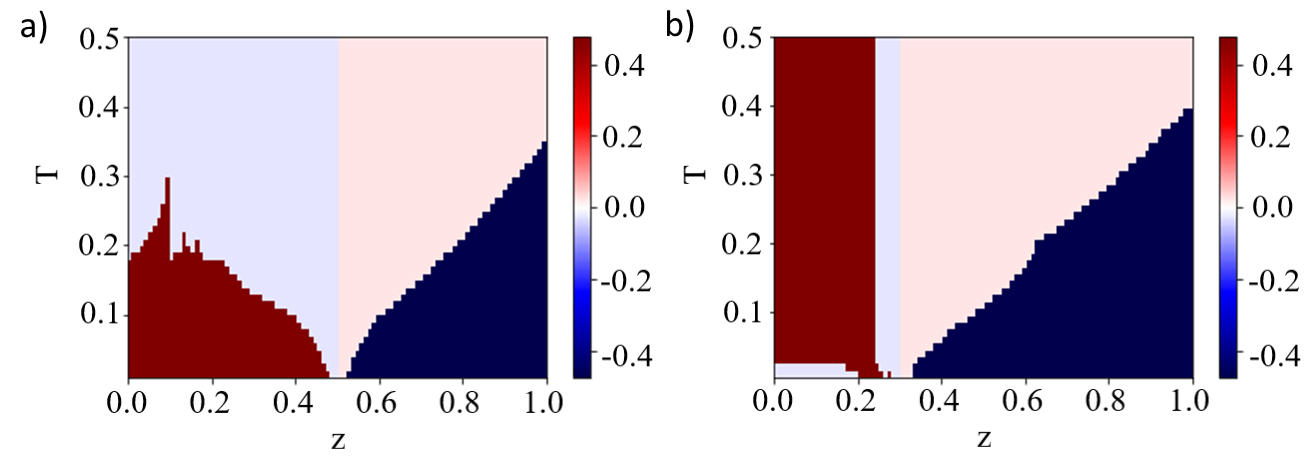}
    \caption{Resta topological phase for thermal ensembles. Topological phase as function of temperature $T$ and next nearest neighbor hopping $z$ for (a) $w=0.5$, $v=0.3$ and (b) $w=0.3$, $v=0.5$. Here is considered a chain of $N=50$ sites. The color code here indicated two three different values for the polarization. Polarization values $P=0$ (white), $P=1/2$ (red) and $P=-1/2$ (blue). Numerical accuracy errors and low impressions of around $10^{-3}$ produce the light regions and have to be taken as $P=0$ (white).}
    \label{fig:phase.png}
\end{figure}

Fig. \ref{fig:phase.png} shows the variation of the Resta topological phase in the ensemble system. As is noted for low temperatures the system has a finite topological phase, however, gradually the regimen of parametric values of hoppings are shrunk, indeed, numerically speaking for $T>0.4$ these topological phases are destroyed everywhere. These values serve as a compass to look for experimental desirable values for hopping modulations. 
\\
However, we have to clarify that the real space representation of the Resta polarization fails in some regimens when $v>w$ because over these cuts the temperature dependence of the chain takes more probability on the edges of the chain, therefore, we have to take attention in this problems that are seen for instance in Fig. \ref{fig:phase.png}(b) where in red for $z<0.3$ it is topological phase however due to the edge states this polarization is not shrunk with temperature due to the dimerized limit, that is not realistic to reach whatsoever. Another mathematical expression for thermal topological phases has to be implemented to have more accurate plots, here for instance the results are more in the qualitative point of view.

\subsection{Optimized Quantum Fisher Information} 
To calculate the QFI consider a complete set of orthonormal states with a density matrix like Eq. 2, in this way, the QFI is defined as \cite{Hauke2016},
\begin{equation}
\begin{split}
\boldsymbol{F}_{Q}[\rho,A]=\frac{1}{2}\sum_{m,n}\frac{(\lambda_m-\lambda_n)^2}{\lambda_m+\lambda_n}|\bra{n}A\ket{m}|^2
\end{split},
\end{equation}
where $\lambda_m$ are the eigenvalues and $A$ is the operator that measures the QFI as an uncertainty relation. The QFI matrix for a qudit-qubit system like the SSH Hamiltonian has a Hilbert space of the form $\mathcal{H}^{d}\otimes\mathcal{H}^{2}$, therefore, using the Pauli matrices $\hat{\sigma}_{i}$, the QFI matrix elements can be written in the following form, 
\begin{equation}
\begin{split}
(M_{site,atom})_{lk}&=\frac{1}{2}\sum_{m,n}\frac{(\lambda_m-\lambda_n)^2}{\lambda_m+\lambda_n}\bra{n}\sigma_{l, atom}\otimes I_{site}\ket{m}\\
&\times\bra{m}\sigma_{k, atom}\otimes I_{site}\ket{n}
\end{split},
\end{equation}
this is a $3\times3$ matrix with each element related to the Pauli matrices selection $l,k=x,y,z$. By a minimization process over the space operator, the OQFI can written as, 
\begin{equation}
\begin{split}
I_P=\min_{\boldsymbol{\sigma}}(M_{site, atom})_{lk}\rightarrow\lambda_{min}
\end{split},
\end{equation}
also known as the interferometric power \cite{Adesso2014}, is proportional to the minimum eigenvalue of the QFI matrix. 
When temperature increments as $T\rightarrow1$ the system reaches a maximally entangled region because this mixture of states becomes equally probable.

The OQFI as a function of temperature $I_{P}(T)$ (Fig. 2(a)) has maximum values in the topological region $v<w$ with $w=0.5$ and $N=40$ unit cells. For $I_{P}(v=0.1)$ the OQFI has more entanglement, which is gradually lost by the increment of the intra-hopping $v$. Conversely, the OQFI as a function of $v$ (Fig. 2(b)) reaches its maximum value in the topological dimerized limit when $v\rightarrow0$ and 

\begin{equation}
I_{P}(v\rightarrow0)\approx 0.6
\end{equation}

Also in the limit of low temperatures, the OQFI presents more entanglement and in the high-temperature regime, thermal effects destroy quantum information in the topological system. In a simple SSH model ($z=0$), topological phase transition $v=w$ is not characterized by the OQFI, and only the topological dimerized limit $v=0$ is captured by this metric. 
\begin{figure}[hbt!]
\centering
    \includegraphics[width=12cm]{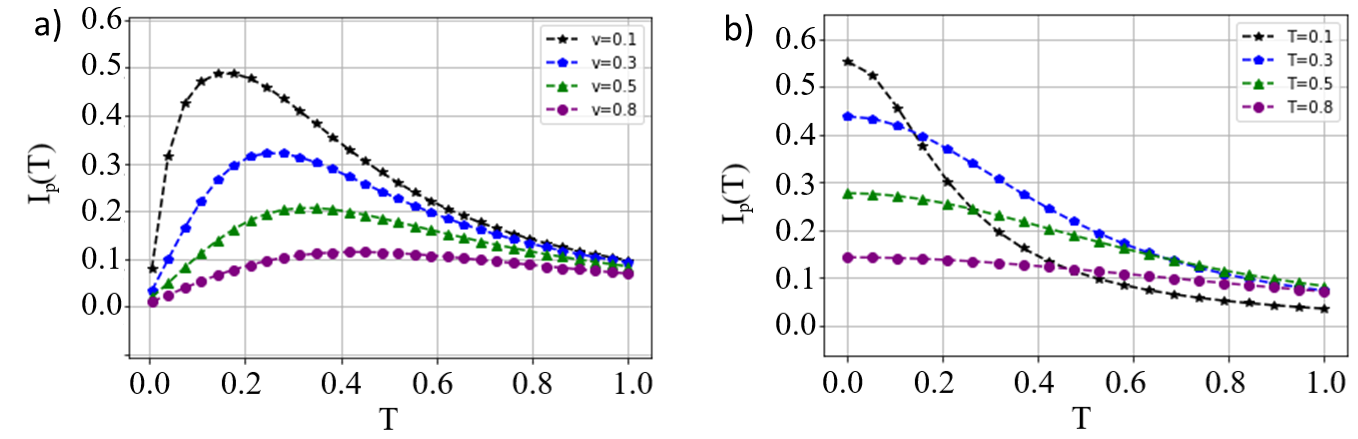}
    \caption{OQFI with parameters $w=0.5$, $z=0$ and $N=50$ unit cells as function of temperature (a) $I_{P}(T)$ and intra-cell hopping (b) $I_{P}(v)$.}
    \label{fig:quantum_fisher_1}
\end{figure}

When second neighbor interaction is taken into account, the OQFI exhibits new features and a richer behavior. The OQFI as a function of temperature (Fig. 3(a),(c)) is similar to the simple case but in the extended SSH model $I_{P}(w=z)=0$ captures the topological phase transition between topological regions with topological phases $\pi\rightarrow-\pi$. The OQFI as a function of the second neighbor hopping $I_{P}(z)$ is shown in Fig. 3(b),(d), where it is more clear that the QFI information goes to zero for topological phase transition $w=z$ in the cases $I_P(z=0.5,v=0.3,w=0.5)=0$  (Fig. 3(b)) and $I_P(z=0.3,v=0.5,w=0.3)=0$ (Fig. 3(d)). Therefore, simple and extended SSH models are quite different in their phase transition because at $v=w$ for $z=0$ the OQFI does not capture this transition, while for $w=z$ in the extended model, the OQFI captures this transition when,

\begin{equation}
I_{P}(w=z)\rightarrow0
\end{equation}

It's important to note that both topological phases $\pi,-\pi$, present a finite OQFI while in the trivial region with phase $0$ goes to zero. These results are purely numerical, further investigation in analytical expressions should be taken to corroborate analytically our findings. \\

\begin{figure}[hbt!]
\centering
    \includegraphics[width=12cm]{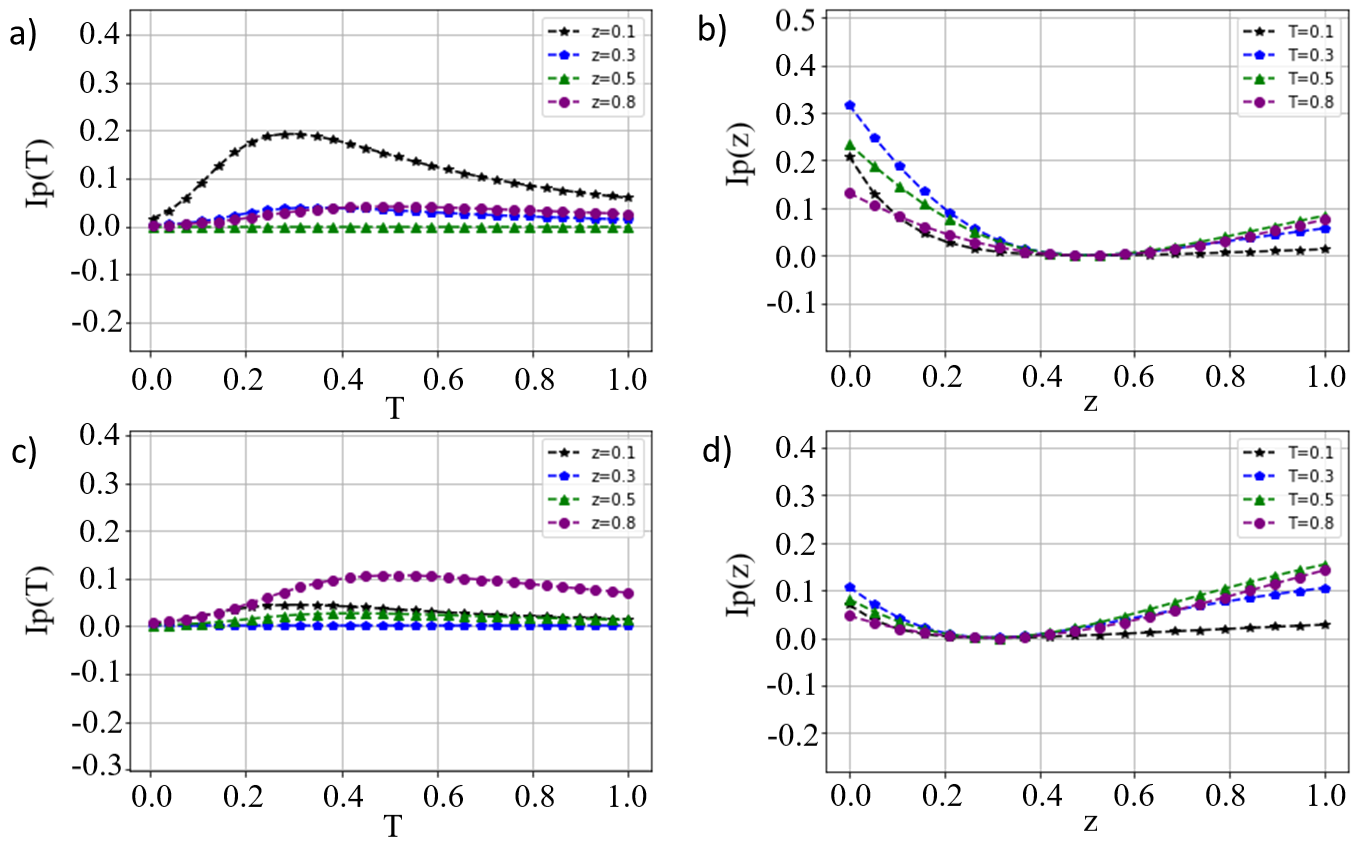}
    \caption{OQFI as function of temperature $T$ and second neighbor hopping $z$ with $N=50$ unit cells. For $v=0.3$, $w=0.5$; (a) $I_{P}(T)$ and (b) $I_{P}(z)$. For $v=0.5$, $w=0.3$; (c) $I_{P}(T)$ and (d) $I_{P}(z)$.}
    \label{fig:quantum_fisher_T}
\end{figure}

\section{Conclusions}
We studied the one-dimensional SSH model with a thermal bath. In the thermal ensemble, the effect of temperature on the topological insulator generates thermal changes in the information, causing it to gradually lose the OQFI; however, due to the presence of topological phases $\pm\pi$, the amount of quantum information was protected up to a finite temperature. The OQFI was not an indicator of the trivial-topological phase transition $0\rightarrow\pi$ but when introducing tunneling to second neighbors $z$ the OQFI characterized the topological-topological phase transition $\pm\pi\rightarrow\mp\pi$. On the other hand, it was determined that there is always an OQFI for the topological phases $\pm\pi$, while, for the trivial phase the OQFI tends to disappear. Further studies in this direction have to be conducted we encourage the study of the OQFI in flat band systems with non-Abelian effects \cite{Navarro-Naumis2021,Navarro-Naumis2024}.

\bibliography{references.bib}
\end{document}